\pdfoutput=1

\documentclass[11pt]{article}

\usepackage{EMNLP2023}

\usepackage{times}
\usepackage{latexsym}

\usepackage[T1]{fontenc}

\usepackage[utf8]{inputenc}

\usepackage{microtype}

\usepackage{inconsolata}

\usepackage{graphicx}
\usepackage{subfig}
\usepackage{appendix}
\usepackage{amsmath}
\usepackage{lipsum}
\usepackage{enumitem}

\usepackage{siunitx}

\begin{document}

\title{Creator Context for Tweet Recommendation}

\author{
Spurthi Amba Hombaiah$^1$~~~~~~Tao Chen$^1$~~~~~~Mingyang Zhang$^1$\\
\textbf{Michael Bendersky$^1$~~~~~~Marc Najork$^2$~~~~~~Matt Colen$^3$}\\
\textbf{Sergey Levi$^1$~~~~~~Vladimir Ofitserov$^3$~~~~~~Tanvir Amin$^3$}
\\ \\
$^1$Google Research \quad $^2$Google DeepMind \quad $^3$Google\\
{\small \texttt{\{spurthiah,taochen,mingyang,bemike,najork,mcolen,sergeyle,vofitserov,tanviramin\}@google.com}}
}


\maketitle

\begin{abstract}
 When discussing a tweet, people usually not only refer to the content it delivers, but also to the person behind the tweet. In other words, grounding the interpretation of the tweet in the context of its creator plays an important role in deciphering the true intent and the importance of the tweet.

In this paper, we attempt to answer the question of how creator context should be used to advance tweet understanding. Specifically, we investigate the usefulness of different types of creator context, and examine different model structures for incorporating creator context in tweet modeling. We evaluate our tweet understanding models on a practical use case -- recommending relevant tweets to news articles. This use case already exists in popular news apps, and can also serve as a useful assistive tool for journalists. We discover that creator context is essential for tweet understanding, and can improve application metrics by a large margin. However, we also observe that not all creator contexts are equal. Creator context can be time sensitive and noisy. Careful creator context selection and deliberate model structure design play an important role in creator context effectiveness. 
\end{abstract}

\section{Introduction}
\label{sec:introduction}

Linguists and philosophers have long recognized the importance of the interplay between utterance semantics and its context~\citep{levinson1983pragmatics}. For instance, the meaning of a statement such as \emph{I am standing here now} can only be interpreted in the context of its speaker. As a more media-related example, news stories often implicitly refer to recent events, assuming common context among their readers (e.g. \emph{Latest news on Ukraine}). While context is important for all media content, its importance is even more pronounced on short-form content platforms like Twitter\footnote{Twitter and tweets were rebranded to ``X'' and ``posts'' respectively in July 2023. However, we use the former names in the paper as this work was carried out prior to July 2023.}. On Twitter, the 140 (and later 280) characters tweet word limits have long inspired idiosyncratic forms of communication~\citep{Westman+Freund:2010}. Therefore, applying computational natural language understanding to tweets alone can be challenging (e.g. \citet{Zampieri2019, Nguyen2020}).
 
 \begin{figure}[t]
    \centering
    \includegraphics[width=0.4\textwidth]{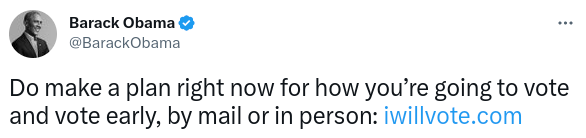}
    \caption{Tweet from Barack Obama}
    \label{fig:barack_obama_tweet}
\end{figure}

\begin{figure}[t]
    \centering
    \includegraphics[width=0.4\textwidth]{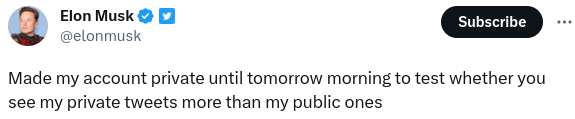}
    \caption{Tweet from Elon Musk}
    \label{fig:elon_musk_tweet}
\end{figure}

 In this paper, we particularly focus on the context of the creator of the tweet as a key to facilitate tweet understanding. Creator context refers to the information about the creator which might not be present in the tweet itself. Take the tweet by Barack Obama in Figure~\ref{fig:barack_obama_tweet} as an example. 
 Without knowing that the creator is a former US President, it is hard to estimate the relationship and the importance of this tweet with regards to the 2020 US presidential election. In another example (Figure~\ref{fig:elon_musk_tweet}), the tweet mentions the pronoun ``my'', however, it is hard to tell who the ``poster'' is, and what event this tweet relates to, from the tweet content alone. Incorporating the creator context can easily address both of these problems. 
 Moreover, knowing 
 the creator is useful for understanding the authoritativeness and news-worthiness of the post, which is highly beneficial for downstream applications like tweet search and recommendation.

Accordingly, in this work, we investigate the importance of creator context for tweet understanding. As Twitter is known as a channel for users to post real-time commentary on world events~\citep{Suarez2018}, we focus on the task of linking tweets to new stories. This task has many practical applications, as news publishers and aggregator services (Google News, Flipboard, Techmeme, Inkl, SmartNews, etc.) often provide Twitter integration in their products. Moreover, this task can assist journalists in composing news articles as they often use tweets as a cited source~\citep{Kapidzic2022}.

The main contributions of this work are as follows:
\vspace*{-5pt}
\begin{itemize}[noitemsep]
 \item We propose a simple yet effective way to mine creator context from an account's metadata.
\item We explore different architectures to incorporate creator context for news-tweet retrieval, and discuss the trade-off between efficiency and effectiveness.
\item We propose a simple yet effective methodology to mine a large scale high-quality corpus of 8M news articles containing embedded tweets, without requiring expensive human annotation for training the models.
\item Our proposed creator context and the retrieval models show strong performance on both the curated dataset and the general tweet stream.
\end{itemize}
\section{Related Work}
\label{sec:related_work}

\textbf{Tweet News Recommendation.} 
Twitter users are known as heavy news consumers~\cite{Kwak2010}.  
To facilitate this, a line of work generates personalized news article recommendation to Twitter users based on their interests (e.g. \citet{Abel2013}). These works often adopt content recommendation techniques, leveraging user's historical posts to build a profile, and comparing it against news articles. Another line of work attempts to perform recommendation at a post level, aiming at linking related news to a specific tweet. The seminal work by ~\citet{Guo2013} uses a graph-based latent variable model to capture tweet-news similarity.

In our work, we study the reverse task of recommending related tweets to news articles. Earlier work by ~\citet{Krestel2015} formulates this as a classification task (relevant or irrelevant). On a human labeled dataset of 340 \emph{$\langle$news, tweet$\rangle$} pairs, they build an AdaBoost model using tweet-news token-level similarity from a document likelihood model, and topic-level similarity from the Latent Dirichlet Allocation model, along with 16 other hand-crafted features such as publication time, tweet length, and follower count of the Twitter user. In another work,~\citet{Suarez2018} build lexical retrieval models to measure the lexical similarity between tweet textual content and query news. They curate and release a human labeled dataset consisting of 100 news articles and 6230 \emph{$\langle$news, tweet$\rangle$} pairs. With this dataset, a recent work by \citet{Thakur2021} directly apply deep retrieval models (most of them trained on the MS-MARCO dataset) in a zero-shot setup, in order to assess the generalization ability of deep models. Unlike the small datasets used by prior works, we collect a large set of 8M \emph{$\langle$news, tweet$\rangle$} positive pairs from news articles with embedded tweets, which enables us to train a deep retrieval model.
Moreover, all the prior works use only the tweet text as opposed to our work of using tweet and creator context together, which brings significant gains as seen in our experiments.

\textbf{Twitter User Modeling.}
Twitter user modeling is often studied in the context of personalized recommendation. Prior works largely leverage the Twitter users' authored posts for user profile modeling. Most of them aggregate the posts and then build user profiles with various semantic granularity, such as token-based,  entity-based, topic-based or category-based (e.g. \citet{Abel2011b,Piao2016}). As tweets are short, some researchers attempt to use external resources, e.g. linked URLs or mined related articles \citep{Abel2011b} to enrich the semantics of tweets. The prior works also find that the user interest is dynamically changing over time \citep{Abel2011a,Piao2016}. Short-term profiles built on recent tweets do not capture the complete user interests well, and all the historical tweets are needed to build a long-term profile~\citep{Piao2016}. This is consistent with our findings (detailed in Section~\ref{sec:creator_context_attributes}). 
However, obtaining all the historical tweets is often technically impractical, especially at large scale and for real-time applications. In our work, we turn to more stable sources of creator context that can be obtained efficiently to approximate the long-term interest.

\section{Methodology}
\label{sec:methodology}

In this section, we first explore potential sources to mine creator context. We then introduce the task of recommending related tweets to news articles, and discuss how the creator context could augment tweet content for this task.

\subsection{Creator Context}
\label{sec:creator_context_attributes}
The metadata of an account is the most accessible information to represent a creator. We list five topically relevant creator attributes that we have explored below. We do not include other creator metadata like the number of followers/followees or the age of the account, as they are not topically related.
\vspace*{-5pt}
\begin{itemize}[noitemsep]
\item Screen handle: The unique identifier of the creator (up to 15 characters).
\item Display name: The full name of the creator as seen on their page (up to 50 characters).
\item Bio: The full text of the creator's bio/profile description (up to 160 characters).
\item Website: The URL of the creator's website (up to 100 characters).
This often encodes creator's affiliation which is helpful to understand this person. 
\item Location: Geographical location of the creator (up to 30 characters). This is a key information to understand tweets about local events.
\end{itemize}

Another important attribute to understand a creator is their previous tweets. However, there are two major challenges to utilize historical tweets for creator modeling. On one hand, a creator could write posts on diverse topics which requires access to all the historical tweets to comprehensively model the topicality \citep{Piao2016}. In practice, it is expensive to obtain a large number of tweets given the Twitter API pricing. On the other hand, creators are actively generating new tweets, and their interests are shifting over time \citep{Abel2011a,Piao2016}; see Appendix~\ref{sec:word_clouds} for examples and further discussion.
This means that, in real world systems, to keep the creator context up to date, ideally we would need to constantly obtain creators' new tweets. This also suggests that, for the task of recommending tweets to news articles, we need access to tweets that could match the time frames of the news articles. 

In our study, we were only able to obtain a few  recent tweets crawled from each Twitter creator's home page. We explored several strategies of utilizing recent tweets in our experiments for creator context modeling, including using all recent tweets, drawing a random sample of recent tweets and using the tweet most similar to the creator bio. However, none of the methods is effective and they are even detrimental to the model performance in news tweet recommendation task.
We posit that this is due to the time mismatch between the recent tweets and news articles. Twitter users' interests shift quickly over time, and thus modeling creators using historical posts requires temporal adaptation to retain performance. This is in line with other works on modeling dynamic content 
\citep{Hombaiah2021, Jin2022}.

In contrast, creators' metadata tend to be stable over time.
In our crawled data, we compared creator metadata from two snapshots which are 3.5 months apart. We found that 90\% of creators have the exact bio (verbatim) and other metadata are also generally static. In the later part of this work, we concentrate on utilizing creator metadata for tweet understanding, for its stability and accessibility.

\begin{figure*}[ht]
    \centering
    \subfloat[\centering Base]{\includegraphics[width=0.2\textwidth]{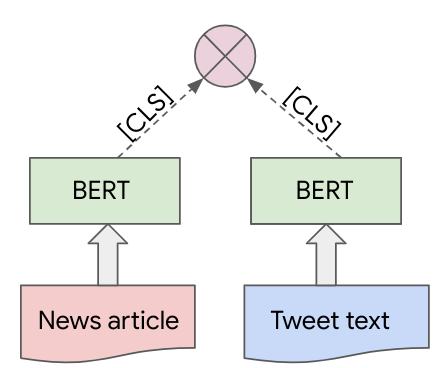}}
    \subfloat[\centering Early Fusion]{\includegraphics[width=0.2\textwidth]{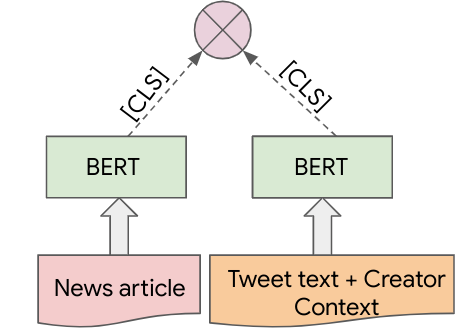}}
    \subfloat[\centering Intermediate Fusion]{\includegraphics[width=0.3\textwidth]{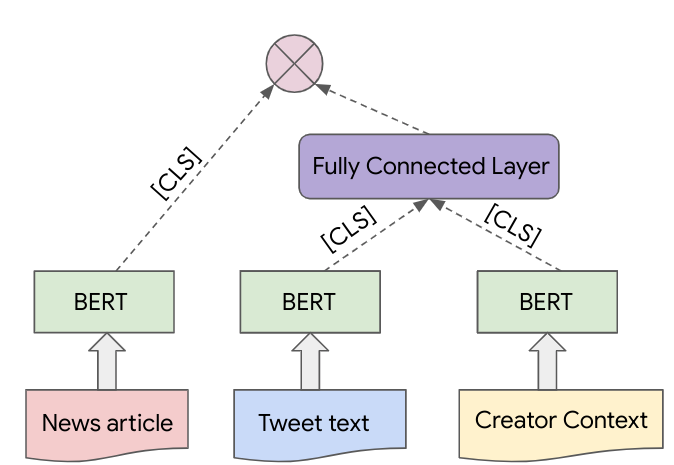}}
    \subfloat[\centering Late Fusion]{\includegraphics[width=0.3\textwidth]{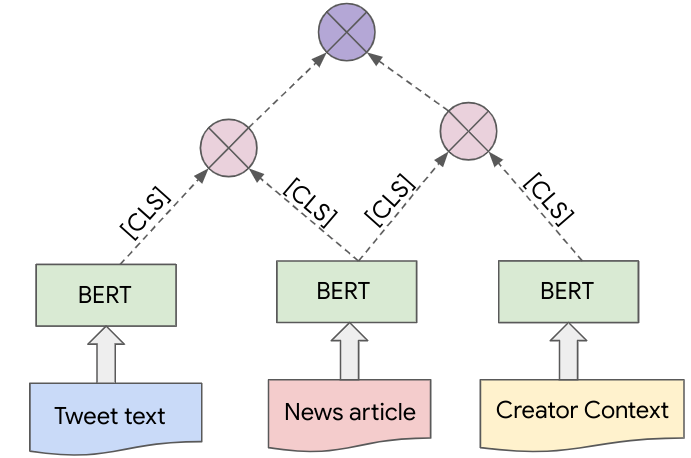}} \\
    \caption{An illustration of different Retrieval Models. A theoretical analysis of the best and worst case complexities of the models can be found in Section~\ref{sec:tweet_recommen}.}
    \label{fig:model_diagrams}
\end{figure*}

\subsection{Recommending Related Tweets to News}
\label{sec:tweet_recommen}

Tweets are a valuable source of real-time news and have been used successfully for news dissemination, and for early detection of breaking events~\cite{weng2011event}. Both news applications and search engines directly surface related tweets for various events and news stories. Moreover, journalists embed tweets in news articles to add depth, authenticity and perspectives to the narrative of their story. Building a model to recommend related tweets to a given news article is an important research task. Such a model could both be beneficial in user-facing applications and as an assistive writing tool for journalists.

We formulate this recommendation task as a retrieval problem: given a news article, we aim at identifying semantically relevant tweets from a large tweet pool. We adopt a dual encoder model for the task as it is a widely used retrieval architecture, and has demonstrated strong performance in many retrieval tasks~\citep{Thakur2021}. Without creator context, the \textbf{Base model} simply consists of two BERT encoders, encoding news article and tweet textual content respectively. The two [CLS] embeddings from the top BERT layer are then used to compute cosine similarity, indicating the semantic relevance between the input news article and the tweet. During serving, we perform nearest neighbor search to obtain top tweets as final results.

Based on the dual encoder framework, we further propose creator context enhanced retrieval models by augmenting tweets with the proposed creator context. For creator context, we combine each attribute with a prefix (``screen'', ``display'', ``bio'', ``website'', ``location'' respectively) as one text sequence. Twitter has length limitations for these attributes and the combined creator context sequences are generally shorter than 391 characters (max length for all attributes and their prefixes put together). Below, we discuss how creator context could be incorporated in different fusion stages (illustrated in Figure~\ref{fig:model_diagrams}). 

\textbf{Early Fusion.}
The most straightforward way to augment tweets is to concatenate tweet text and creator context as one single input sequence for the tweet encoder. This allows the powerful cross-attention mechanism of BERT to model the interaction between a tweet and its creator context and generate a creator-aware tweet embedding. 
However, the time complexity of the attention mechanism is quadratic to the size of its input, making early fusion computationally expensive since the combined input is much longer than the tweet alone.

\textbf{Intermediate Fusion.}
Unlike the early fusion, we use a separate BERT encoder to learn creator context embedding. We concatenate the two [CLS] embeddings from the tweet and the creator encoders and feed them to a fully connected layer, which generates the final creator-aware tweet embedding. During model serving, the creator embedding is computed on a per-creator basis, and could potentially be pre-computed. Compared to early fusion, this helps to significantly reduce computational cost, as embeddings of popular creators only need to be computed once. 

\textbf{Late Fusion.} We further push the fusion to a later stage. We directly use the [CLS] embedding from the creator encoder to compute a \emph{$\langle$news, creator$\rangle$} cosine similarity score, and linearly combine it with a \emph{$\langle$news, tweet$\rangle$} cosine similarity score with a weight.
We first tried to learn this weight as a model parameter. However, the learned weight is biased towards the creator encoder and did not work well on the development set. This is due to the lack of hard negatives in training (we use in-batch negatives) and the creator context alone becomes a distinctive feature. Instead, we consider this weight as a hyperparameter, and tune it on the development set via grid search (detailed in Section~\ref{sec:retrieval_results}).
Similar to the intermediate fusion, this model can reduce the computational cost of creator embeddings. The tuned weight can also provide better interpretability to indicate the contribution from the tweet and creator encoders. However, this model requires two retrievals and a score combining procedure, and thus is less efficient in serving.

\textbf{Model Complexity.} We use $n$ and $m$ to denote the tweet and creator context lengths, respectively. The best and worst case complexities for the Base model is $O(n^2)$ and the Early Fusion model is $O((n+m)^2)$. For Intermediate and Late Fusion models, the worst case complexity is $O(n^2 + m^2)$ when the system is just deployed and the best case complexity is $O(n^2)$ when the system has pre-computed embeddings for all the creators after the system has been running for a sufficient period of time. Among our proposed models, the most efficient ones are the Intermediate and Late Fusion models (excluding the Base model).

\section{Experiments \& Analysis}
\label{sec:experiments}
In this section, we evaluate our creator context enhanced retrieval models for recommending related tweets to a news article. We compare the performance of different types of creator context and different model structures for incorporating it.

\subsection{Dataset}
\label{sec:dataset_preprocessing}

There are two prior works which curated very small news-tweet datasets via human annotation. ~\citet{Krestel2015} selected 17 news articles and annotated 20 tweets (from top results of their model) per article, and ~\citet{Suarez2018} selected 100 news articles derived from Signal1M~\citep{Signal1M2016} and collected human ratings for 62 tweets on average per article. Both datasets are too small to train a deep retrieval model.

Unlike prior work relying on expensive human annotation, we collect positive labels from existing news articles which have embedded tweets published during 2006 -- 2022. Those tweets are carefully selected by the journalists when composing the articles. To be more specific, we obtained a large crawled news article dataset (similar to ~\citet{Liu2021}) and filtered out articles that do not have any embedded tweets or have more than 20 (likely spammy). 
We crawled the metadata from the Twitter creators' profile pages, including the screen handle ($91\%$ coverage), display name ($91\%$), bio ($87\%$),  website ($73\%$) and location ($69\%$).
In total, we collected ~8M \emph{$\langle$news article, embedded tweet$\rangle$} pairs, of which there are 5.3M unique news articles (see Appendix \ref{sec:dataset_ex} for some examples). 

To further validate the relevance of the embedded tweets, we performed a small scale annotation. We sampled 100 news articles along with the top five tweets retrieved by our best performing retrieval model and one tweet originally embedded in the article (if it was not retrieved). Four annotators (authors of this paper) were asked to select the most relevant tweet to the article in question among these tweets (in shuffled order). From the annotation results, we see that for $90\%$ of cases, the original embedded tweet is selected as the most relevant one. This study, albeit at a small scale, demonstrates the validity of using embedded tweets as a surrogate for relevance.

\subsection{Experimental Setup}
For all the models, we use a 12-layer BERT base model ~\cite{Devlin2019} for each encoder, which is initialized using an ``i18n'' checkpoint pre-trained on a large news dataset ~\cite{Liu2021}. We also adopt their vocabulary which consists of $\sim 500$K wordpiece tokens. Following~\citet{Liu2021}, we concatenate news title and body text (additionally remove embedded tweets to avoid information leak), and truncate the content over 512 wordpiece tokens. We allow up to 128 wordpiece tokens for both tweet and creator encoders, as tweet text (up to 280 characters) and creator context are short (up to 391 characters over all the contextual attributes and their prefixes). In the early fusion model, the max sequence length for the combined tweet and creator context encoder is 256 wordpiece tokens. We split \emph{$\langle$news, embedded tweet$\rangle$} pairs for training / development / testing in a ratio of 8:1:1.

We use the Wordpiece tokenizer \citep{Wu2016} to tokenize articles. For tweets and creator context, we first extract intact tokens from hashtags and user mentions based on the following two rules (similar to \citet{Hombaiah2021}):
\begin{itemize}[noitemsep]
\vspace*{-6pt}
\item Split by camelcase and underscore whenever possible.
\item If the above doesn't work, segment the compound word using a dictionary of unigrams constructed from Google n-grams\footnote{\url{https://console.cloud.google.com/marketplace/product/bigquery-public-data/google-books-ngrams-2020}} such that the probability of segmentation is maximized (by assuming conditional independence between the different segments).
\end{itemize}

 As our dataset only contains positive pairs, we use in-batch negatives for model training.  We optimize the model with sigmoid cross entropy loss and in-batch loss. We froze the parameters of the news encoder, as this pre-trained checkpoint has been well trained on news documents, while fine-tuning parameters for other encoders. For all the models, we carefully tuned hyperparameters including learning rate, batch size and training steps on the development set. We used a batch size of 256 and a learning rate of 1e-5 for training our models. All models were trained for 200k steps. For other hyperparameters, we use the same values as those used for training the standard BERT model~\citep{Devlin2019}. 

As this is a retrieval task, we primarily focus on the recall metric (recall@1K), but also report precision at top positions (1 and 5) and Mean Reciprocal Rank (MRR). We retrieve 1k tweets for every news article from the test set tweets for evaluation. For the Late Fusion model, we first retrieve the top 20k tweets using tweet and creator embeddings separately for each article. Then, we compute a combined cosine similarity score using a weight (chosen via grid search on the development set), and re-rank tweets based on the combined score.

For comparison, we adopt the Terrier implementation~\cite{Ounis+2005Terrier} of BM25, a classical lexical retrieval model, as a baseline.
For news-tweet lexical retrieval, ~\citet{Signal1M2016} found that using article title as query has comparable performance to using a longer query from article body. 
Hence, we use only news article titles as queries and retrieve 1k tweets for each article.

\subsection{Retrieval Results}
\label{sec:retrieval_results}

\begin{table}[th]
\begin{center}
\caption{Tweet recommendation results. *, \#, †, ‡ indicate statistically significant (via paired \textit{t}-test with \textit{p} < 0.05) improvements over ``BM25'', ``Base'', ``Late Fusion'' and ``Intermediate Fusion'' models respectively.}
\label{table:retrieval_model_results}
\scriptsize
\begin{tabular}{{|c|S[table-format=1.3]|S[table-format=1.3]|S[table-format=1.3]|S[table-format=1.3]|S[table-format=1.3]|}}
\hline
\textbf{Model} & \textbf{~~~~P@1~~~~} & \textbf{~~~~P@5~~~~} & \textbf{~~R@1000~~}  & \textbf{~~~~~MRR~~~~~}\\
\hline
BM25 & 0.116 & 0.181 & 0.439 & 0.153\\
\hline
\hline
Base & 0.311\textsuperscript{*} & 0.461\textsuperscript{*} & 0.801\textsuperscript{*}  & 0.391\textsuperscript{*}\\
\hline
\hline
Early & 0.362\textsuperscript{*\#†‡}	& 0.527\textsuperscript{*\#†‡} & 0.875\textsuperscript{*\#†‡} & 0.449\textsuperscript{*\#†‡}\\
\hline
Intermediate & 0.354\textsuperscript{*\#†} & 0.519\textsuperscript{*\#†} & 0.871\textsuperscript{*\#†}  & 0.441\textsuperscript{*\#†}\\
\hline
Late & 0.228\textsuperscript{*} & 0.379\textsuperscript{*} & 0.814\textsuperscript{*\#} & 0.309\textsuperscript{*}\\
\hline
\end{tabular}
\end{center}
\end{table}

Table~\ref{table:retrieval_model_results} shows the results for the five models on the test set. The deep retrieval models perform significantly better than the lexical retrieval model BM25. Overall, the proposed models where we use creator context perform better than the Base model for which the creator context is not used, especially when considering recall. This proves the effectiveness of creator context for tweet modeling. Out of the three models which use creator context, the Early Fusion model performs the best, significantly improving the Base model performance by a relative measure of $9.2\%$ on Recall@1K, $16.4\%$ on Precision@1 and $14.8\%$ on MRR. The Intermediate Fusion model is a close second. The additional improvement of the Early Fusion model is probably from the lower layer interactions between tweet and creator context. On the other hand, the Intermediate Fusion model, though slightly less effective, is much more efficient than Early Fusion as it allows using pre-computed creator embeddings. This could lead to significant cost and latency savings, especially for large scale user facing applications. 

The Late Fusion model performs the worst among the three models using creator context. Compared to the Base model, its recall is better but other ranking focused metrics are worse. The improvement on recall is again an indication of the usefulness of creator context, but the low precision is a strong indication that a weighted sum of similarity scores fails to capture the needed interactions between tweet and creator context.

\textbf{Model Recommendation.} Based on the results, we recommend using the \textbf{Early Fusion} model if optimal effectiveness is desired. However, for an industry setup where serving cost and latency are mission-critical, using the \textbf{Intermediate Fusion} model would be highly beneficial. The Intermediate Fusion model allows pre-computation of creator embeddings unlike the Early Fusion model where they are recomputed 
for each tweet. As an additional practical benefit, decoupling computation of creator embeddings from tweet embeddings can enable them to be used separately in other applications like predicting creator similarities, matching creators to queries etc.

\subsection{Creator Context Importance Analysis}
\label{sec:creator_context_attributes_importance_analysis}
We also investigate the importance of different creator attributes (discussed in Section~\ref{sec:creator_context_attributes}) for modeling tweets. We conduct an ablation experiment by leaving out one attribute at a time, training our best performing Early Fusion model on the training set, and reporting its average loss on the test set.

As we see from Table~\ref{table:importance_creator_context}, creator bio is the most useful creator context, as ablating it leads to the largest increase in loss. Creators often mention their interests and professions in their bio. This contextual information helps models to better understand tweets. Other creator context types are not as helpful as creator's bio, but they still bring benefits to the task of recommending tweets.

\vspace*{-5pt}
\begin{table}[th]
\begin{center}
\caption{Importance of different creator context. Lower loss denotes better performance.}
\label{table:importance_creator_context}
\small
\begin{tabular}{{|l|c|}}
\hline
\textbf{Creator Attribute} & \textbf{Loss}\\
\hline
All & $0.064$\\
\hline
\hspace{0.3cm} w/o Screen handle & $0.067$\\
\hline
\hspace{0.3cm} w/o Display name & $0.068$\\
\hline
\hspace{0.3cm} w/o Bio & $0.077$\\
\hline
\hspace{0.3cm} w/o Website & $0.065$\\
\hline
\hspace{0.3cm} w/o Location & $0.066$\\
\hline
\hline
No author context & $0.100$\\
\hline
\end{tabular}
\end{center}
\vspace{-3mm}
\end{table}
\section{Discussion}
\label{sec:discussion}

Our dataset is curated from news articles and their embedded tweets. Those tweets are chosen by journalists and might be biased towards certain creators (e.g. popular figures/celebrities). We thus explore whether our model can generalize to the general population of tweets from arbitrary creators and if creator context can be useful for modeling tweets about local, rare and special-interest events.

To this end, we obtain a large set of public tweets from the Internet Archive (detailed in Appendix~\ref{sec:internet_archive}). As news events are highly time sensitive, for each article, we restrict its candidate tweets to all tweets posted no earlier than one week of the article posting time (unlike the experiments in Section~\ref{sec:experiments} which used all tweets without any time restriction as retrieval candidates). We use our best performing Early Fusion model to retrieve relevant tweets and perform a small scale annotation for a quality check. To be specific, we randomly pick three dates (07/31/2017, 02/27/2018 and 06/24/2019), and then randomly sample 100 articles published on the three dates. We (three authors of this paper) annotated the relevance of top five retrieved tweets for each article. We find that our model achieves a high precision -- 91\% of articles have at least one relevant tweet in the top five results. This demonstrates that our model could perform well on the general tweet population. Please see Appendix~\ref{sec:model_gen_ex} for examples. 

We also find that the creators of the top retrieved results are diverse, consisting not only of popular figures but also ordinary people like local residents. See Figure~\ref{fig:news_article_top_tweet_ex_2} for an example, where, for a \href{https://www.channel3000.com/features/olympic-gold-medalist-matt-hamilton-returns-to-wisconsin/article_bb055482-99ee-5f12-9cfd-aa3a7af7310a.html}{news article} about an Olympic athlete returning to his home ground Wisconsin, the top retrieved tweet comes from a local resident. The creator's Twitter page mentions ``Kimberley, Wisconsin'' as the geo location making the creator context very useful in this context.

Moreover, we observe that creator understanding is particularly useful for modeling tweets about local and less popular events.
Figure~\ref{fig:news_article_top_tweet_ex_3} shows an example of the top tweet retrieved by our model (which is also embedded) for a \href{https://www.ctinsider.com/california-wildfires/article/Vegetation-fire-breaks-out-in-San-Jose-14037622.php}{news article} about a local news of a vegetation fire in San José, California. The creator's display name ``San José Fire Dept.'' and their geo location ``San José, California'' are critical in matching and recommending this particular tweet for the news article.

\section{Conclusion \& Future Work}
\label{sec:conclusion_future_work}
In this paper, we investigated the problem of using creator context for tweet understanding.  Different types of creator context, including creators' bio, screen handle, display name, website and location are explored. We also proposed and examined three different model structures to incorporate creator context for the task of recommending tweets to news articles. We demonstrated that, with creator context, significant quality improvements can be achieved. We also showed that not all creator contexts are equal, and they have different effectiveness. For example, creators' tweets, as their topics shift quickly, when used as creator context, can adversely affect performance, especially when the tweets and news articles are not temporally aligned. 

As future work, we would like to explore how non-topical creator features like followers count and creator social graph could be incorporated into tweet recommendation. We also plan to investigate the usefulness of creator context for other tasks (e.g., event detection) and other platforms (e.g., TikTok, YouTube Shorts).

\section*{Ethical Considerations}
\label{sec:ethics}
To the best of our knowledge, our work is ethical and has a positive impact on society and well-being of humans. We treat our data with utmost care and confidentiality. The dataset of news articles, tweets and creator information we use is encrypted and access protected. We periodically update the data to ensure we do not use any stale creator information which can compromise what creators want to reveal about themselves in the public.

In this paper, we focused on the usefulness of creator context but did not check if creator context (which is usually self-reported) is reliably truthful. Some form of creator context quality assurance may be warranted when using the proposed methods in real-world systems, where creators may be incentivized to ``game the system'' by inflating their biography. 

\bibliographystyle{acl_natbib}
\bibliography{acl}

\clearpage
\appendix
\newpage
\begin{appendix}
\section{Shift in Creator Context Tweets Over Time}
\label{sec:word_clouds}
\begin{figure}[ht]
    \centering
    \subfloat[\centering Nov 2022]{\includegraphics[width=0.25\textwidth]{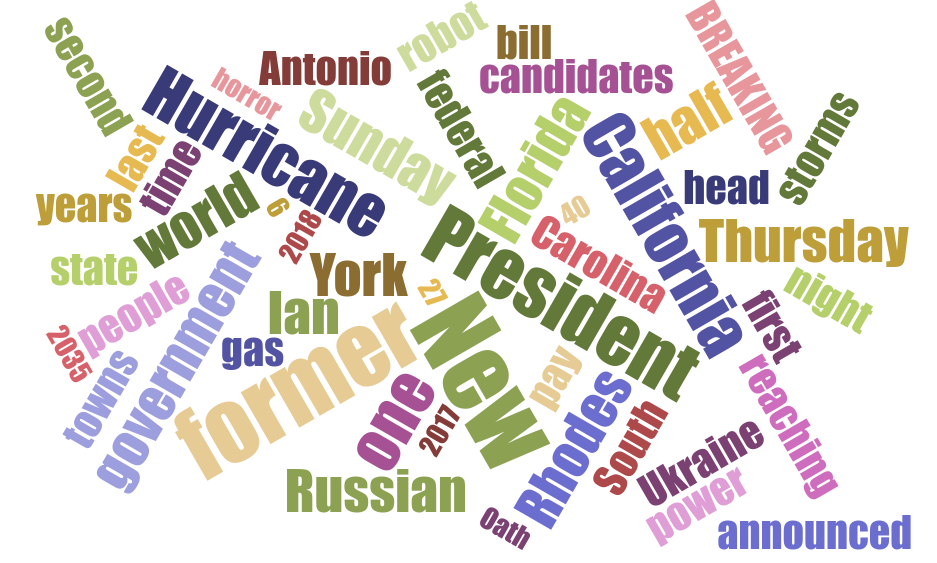}}
    \subfloat[\centering Feb 2023]{\includegraphics[width=0.25\textwidth]{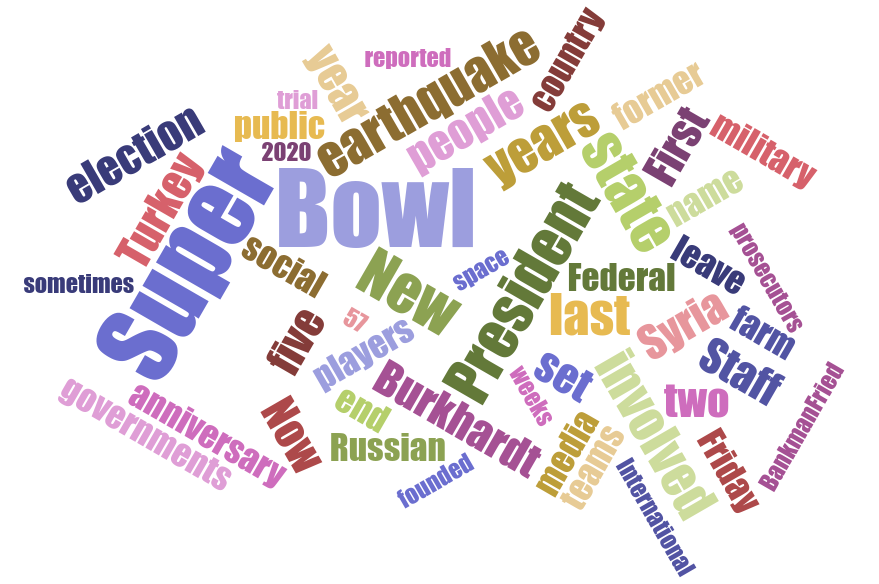}} \\
    \caption{Word clouds for Associated Press.}
    \label{fig:ap}
\end{figure}

\begin{figure}[ht]
    \centering
    \subfloat[\centering Nov 2022]{\includegraphics[width=0.25\textwidth]{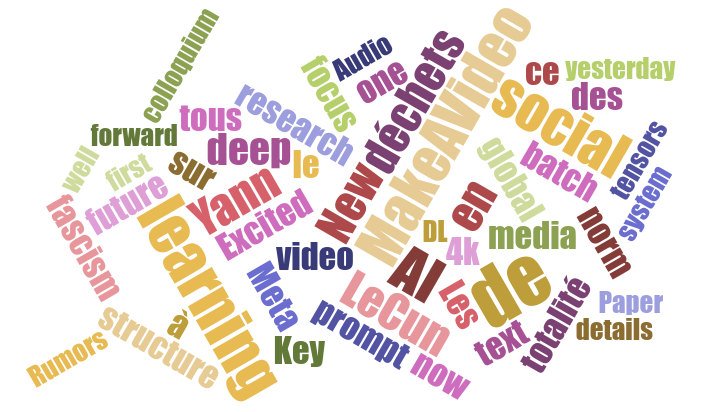}}
    \subfloat[\centering Feb 2023]{\includegraphics[width=0.25\textwidth]{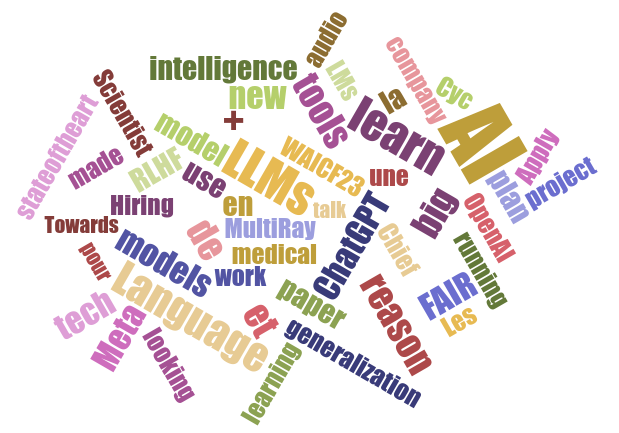}} \\
    \caption{Word clouds for Yann LeCun.}
    \label{fig:yann}
\end{figure}

To illustrate the shift in creators' interests and the topics they tweet about, we show word clouds for two Twitter creators, Associated Press (\url{https://twitter.com/AP}) and Yann LeCun (\url{https://twitter.com/ylecun}) over a sample of recent tweets from the respective creators during two different time periods, November 2022 and February 2023. From Figure ~\ref{fig:ap}, for Associated Press, we see that some of the main topics during Nov'22 were around ``Hurrican Ian'', ``Russia-Ukraine war'' etc. However, the main topics are around ``Turkey \& Syria earthquake'', ``Super Bowl'' etc during Feb'23. For the Twitter creator, Yann LeCun, from Figure ~\ref{fig:yann}, though most important topics largely concern deep learning in both time periods, trending topics ``LLMs'' and ``ChatGPT'' make an appearance in Feb'23.

\vspace{-9mm}

\vspace{-3mm}

\section{Dataset Examples}
\label{sec:dataset_ex}
\begin{figure}[th]
    \centering
    \includegraphics[width=0.48\textwidth]{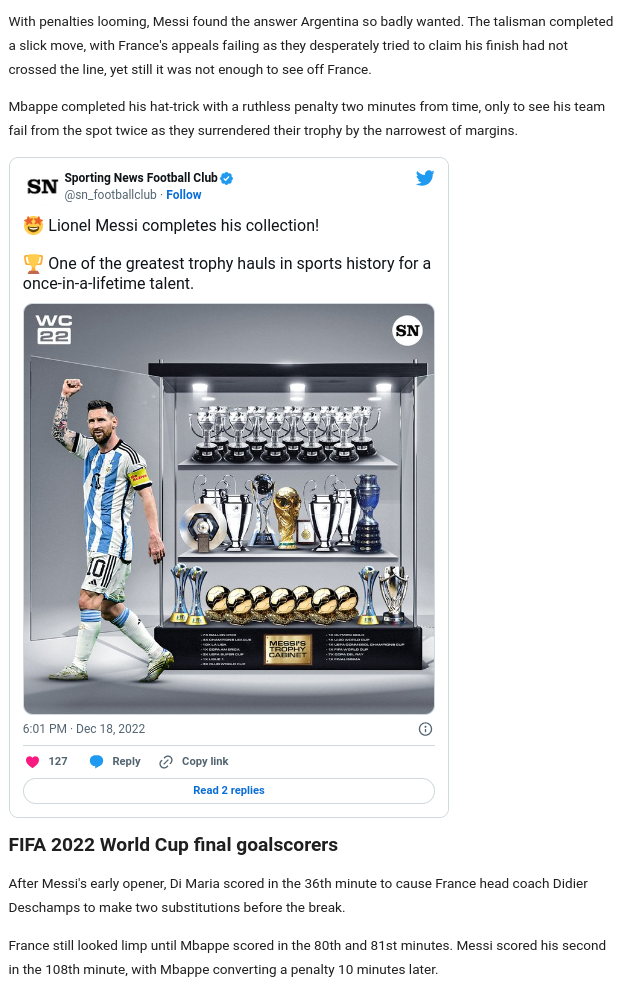}
    \caption{Snippet of a news article in English titled ``Who won the 2022 FIFA World Cup? Final score, result and highlights from Qatar title decider'' with an embedded tweet in English. URL:  \href{https://www.sportingnews.com/uk/football/news/who-won-2022-fifa-world-cup-final-score-result-highlights-qatar/aar5gfviuuapvkmcq3tf7dep}{https://www.sportingnews.com/uk/football/news/who-won-2022-fifa-world-cup-final-score-result-highlights-qatar/aar5gfviuuapvkmcq3tf7dep}.}
    \label{fig:news_article_embedded_tweet_messi}
\end{figure}

Figure~\ref{fig:news_article_embedded_tweet_messi} shows an example of a snippet of a news article in English about the FIFA 2022 World Cup with an embedded tweet about Messi's trophies over the years.

\begin{figure}[th]
    \centering
    \includegraphics[width=0.5\textwidth]{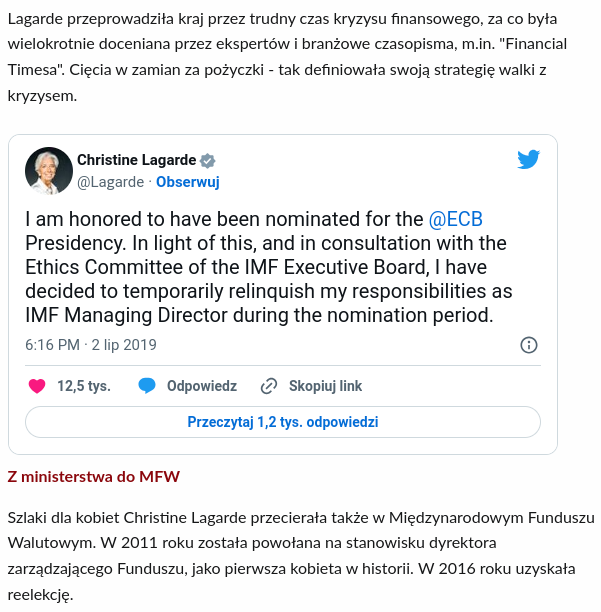}
    \caption{Snippet of a news article in Polish titled ``Christine Lagarde nową szefową Europejskiego Banku Centralnego. Kim jest?'' with an embedded tweet in English. URL: \href{https://polskieradio24.pl/42/273/Artykul/2334907}{https://polskieradio24.pl/42/273/Artykul/2334907}.}
    \label{fig:news_article_embedded_tweet_christine}
\vspace{-3mm}
\end{figure}

Figure~\ref{fig:news_article_embedded_tweet_christine} shows an example of a snippet of a news article in Polish about the French politician Christine Lagarde's appointment as the President of the European Central Bank with a related embedded tweet from Lagarde in English. This example also illustrates that our model is trained to capture cross-lingual relations.

\begin{figure}[h]
    \centering
     \subfloat[\centering Embedded tweet]{\includegraphics[width=0.48\textwidth]{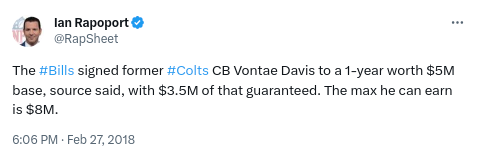}} \\
     \subfloat[\centering Retrieved tweet]{\includegraphics[width=0.48\textwidth]{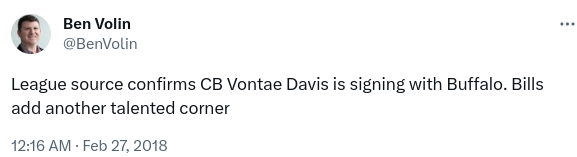}}
    \caption{Article original embedded tweet (chosen by the journalist) and top retrieved tweet by our model for a news article titled ``Bills sign former Colts CB Vontae Davis''. URL: \href{https://coltswire.usatoday.com/2018/02/27/buffalo-bills-sign-indianapolis-colts-vontae-davis/}{https://coltswire.usatoday.com/2018/02/27/buffalo-bills-sign-indianapolis-colts-vontae-davis}.}
    \label{fig:news_article_top_tweet_ex_1}
\vspace{-3mm}
\end{figure}

\section{Internet Archive Dataset}
\label{sec:internet_archive}
To construct a tweet dataset to verify model generalization, we use the public crawl of tweets from the Internet Archive\footnote{\url{https://archive.org/details/twitterstream}}. As we sample articles from three dates (07/31/2017, 02/27/2018 and 06/24/2019), we collect tweets posted no earlier than one week for the corresponding date (i.e., 07/24/2017 - 07/30/2017, 02/19/2018 - 02/26/2018 and 06/17/2019 - 06/23/2019). This results in $4.9$M, $5.5$M and $4.8$M unique original tweets (no retweets) respectively. We extract the tweet text and the creator context information from the tweets for inference. Each week of tweets are used as retrieval candidates for their corresponding articles.

\section{Model Generalization Examples}
\label{sec:model_gen_ex}
We show a top retrieved tweet by our model from the Internet Archive dataset, along with the article and its original embedded article (chosen by the journalist who composed the article).  In Figure~\ref{fig:news_article_top_tweet_ex_1}, the news article is about the ``Buffalo Bills signing a former Indianapolis Colts cornerback Vontae Davis'' and our top retrieved tweet is highly topically related to the news article. 

\begin{figure}[!htbp]
\vspace{-44mm}
    \centering
     \subfloat[\centering Embedded tweet]{\includegraphics[width=0.5\textwidth]{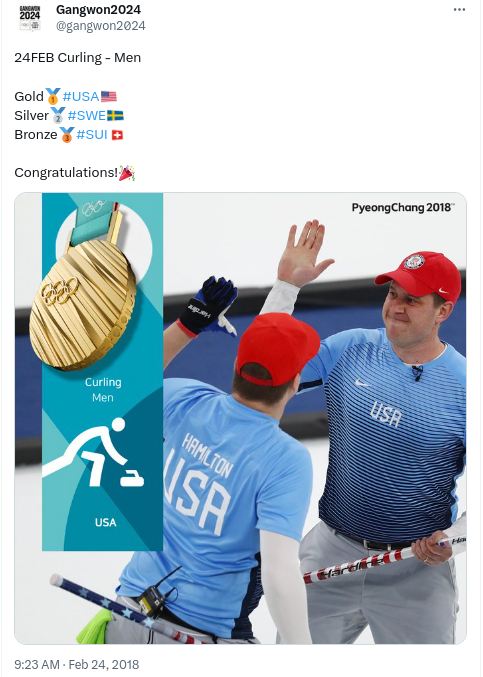}}\\
     \subfloat[\centering Retrieved tweet]{\includegraphics[width=0.5\textwidth]{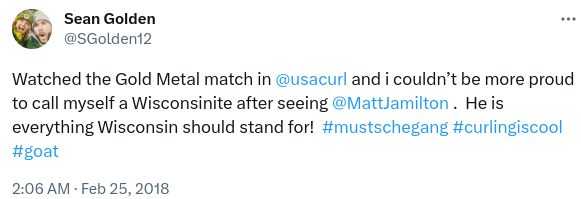}}
    \caption{Article original embedded tweet (chosen by the journalist) and top retrieved tweet by our model for a news article titled ``Olympic gold medalist Matt Hamilton returns to Wisconsin''. URL: \href{https://www.channel3000.com/features/olympic-gold-medalist-matt-hamilton-returns-to-wisconsin/article_bb055482-99ee-5f12-9cfd-aa3a7af7310a.html}{https://www.channel3000.com/features/olympic-gold-medalist-matt-hamilton-returns-to-wisconsin/article\_bb055482-99ee-5f12-9cfd-aa3a7af7310a.html}.}
    \label{fig:news_article_top_tweet_ex_2}
\end{figure}

Figure~\ref{fig:news_article_top_tweet_ex_2} shows an article titled ``Olympic gold medalist Matt Hamilton returns to Wisconsin''.
One of the top retrieved tweets by our model is from a local Wisconsin resident (a non-celebrity; the creator's Twitter page mentions ``Kimberley, Wisconsin'' as their location) and very topical as the article concerns a Wisconsin athlete. While the original embedded tweet is about the medals won during the ``Curling - Men's event'' (posted by the official account of Gangwon 2024 Winter Youth Olympic Games), our retrieved tweet not only is more relevant to the overall topic of the article, but also offers perspectives from a local resident. This demonstrates that our models, although having been trained on tweets embedded in news articles, generalize well over the general tweet population.

\begin{figure}[!htbp]
\vspace{-0.4mm}
    \centering
     \subfloat[\centering Embedded and Retrieved tweet]{\includegraphics[width=0.5\textwidth]{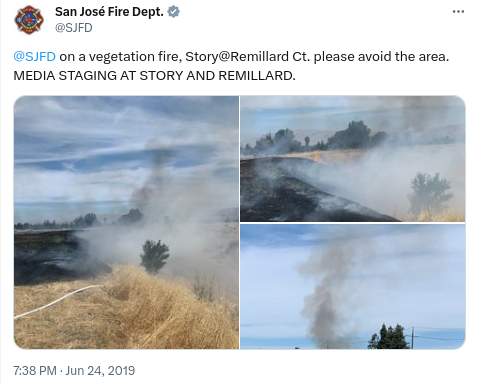}}
    \caption{Article original embedded tweet (chosen by the journalist) and top retrieved tweet by our model for a news article titled ``Vegetation fire contained in San Jose''. URL: \href{https://www.ctinsider.com/california-wildfires/article/Vegetation-fire-breaks-out-in-San-Jose-14037622.php}{https://www.ctinsider.com/california-wildfires/article/Vegetation-fire-breaks-out-in-San-Jose-14037622.php}.}
    \label{fig:news_article_top_tweet_ex_3}
\end{figure}

Figure~\ref{fig:news_article_top_tweet_ex_3} shows a local news article titled ``Vegetation fire contained in San Jose'' about a vegetation fire in San José, California . The top tweet retrieved by our model which is also the embedded tweet in the article is from the official account of the ``San José Fire Department''. The creator's display name, ``San José Fire Dept.'' and their location ``San José, California'' are particularly useful as creator context. This demonstrates that creator context can be useful for modeling tweets for rare and local events.

\vspace{-400mm}

\end{appendix}

\end{document}